# Correlation between Content and Traffic of the Universities' Website

**Bahram Kalhor**
Department of Computer, College of Mechatronic, Karaj Branch, Islamic Azad University, Alborz, Iran
Kalhor_bahram@yahoo.com

**Alireza Nikravanshalmani**
Department of Computer, College of Mechatronic, Karaj Branch, Islamic Azad University, Alborz, Iran.
nikravan@kiau.ac.ir

**Abstract**
The purpose of this study is to analyse the correlation between content and traffic of 21,485 academic websites (universities and research institutes). The achieved result is used as an indicator which shows the performance of the websites for attracting more visitors. This inspires a best practice for developing new websites or promoting the traffic of the existing websites. At the first step, content of the site is divided into three major items which are: Size, Papers and Rich Files. Then, the Spearman correlation between traffic of the websites and these items are calculated for each country and for the world, respectively. At the next step, countries are ranked based on their correlations, also a new indicator is proposed from combining these three correlations of the countries. Results show that in most countries, correlation between traffic of the websites and Papers is less than correlations between traffic of the websites and Rich Files and Size.

**Keywords:** Rankings of countries, Performance of site, Traffic, Content of site, Papers, Size of site

## Introduction

Traffic of a site is amount of visitors and visits a site receives. The relationships between traffic and quality of the websites have not been considered as much as the relationships between Web hyperlink data and such performance measures. There is a significant correlation between web traffic and academic quality (Vaughan & Yang, 2013). Finding the correlation between the traffic and content's quality of an academic website can be used to develop a new or to promote an existing website. Traffic of the websites is used as an indicator by some ranking systems such as 4icu.org and websanji.ricest.ac.ir to rank universities.

To our best knowledge, the correlation between the traffic of the websites and their major partitions which are: Size, Papers and Rich Files have not yet been the focus of considerable study. Comparing the correlations between traffic of a site and items such as: number of the papers (Papers), number of the content files (provided in pdf, doc and ppt formats referred to as Rich Files) and the total number of files and pages included in the site (Size) can give useful information regarding visitors' behavior and their interests. These relationships can help the designer to find which item is useful to be used for development of websites in the future. In particular, we focus on the calculating correlation between traffic of the websites and the major partitions of the academic websites for more than 21,485 websites. We investigate the performance of the each data structure for obtaining more traffic.

In this study, the first step for investigating relation between traffic of the websites and the major parts of the websites is choosing trusted data sources. Alexa is reliable for collecting traffic data (Vaughan & Yang, 2013) which provides rankings of the websites based on number of the visitors and hits. Second step is choosing suitable site explorer or search engine for extracting data. Google have been chosen for collecting size of the websites and number of the files for each university based on (Aguillo, et al, 2006). We use the Google Scholar for collecting number of the papers which has been used in many rankings. The final step is dividing websites into regular partitions. Content of the websites are divided into three major partitions which are: size of the site, number of the papers and number of the files. The organization of this paper is as follows: at next section the literature associated to the research topic has been presented, and then the methodology of the proposed system is discussed. At next section the results of the research has been demonstrated. Limitations and discussions followed by conclusion and future work of discussed at the final sections of the paper.

## Literature Review

Researches in the field of information sciences to find the visitors' behavior of websites have received the considerable attention, recently. In the study carried out by (Vaughan & Yang, 2013), groups of universities and businesses are selected from the United State of America and China. They have used and compared the traffic data of three sources: Alexa, Compete and Google Trend. They found a significant correlation between traffic of the websites and academic quality performance. Also they found that the Alexa is better than two other data sources. They suggested that using data from one source, preferably Alexa, is likely to be sufficient for practical purposes. They mentioned that the limitation of their study is covering two countries.

Unfortunately, websites' ranks in the rankings of universities have been their basic parameter for calculating correlation with traffic of the websites. Correlations between traffic of the websites and real data have not been considered in their study.

They also didn't investigate the similarity between traffics of the domains and subdomains in the Alexa. Although Google and Google Scholar provide sub report for subdomains, respectively, the Alexa assigns traffic of the main domain to its subdomains. In this case, correlations between the traffic of the main domains are reliable, but if site of the university or institute has been located in the subdomain the correlation between traffic of the subdomain and data of subdomain isn't reliable. In United State of America we found many websites of this type which have been located in the subdomains.

(Aguillo, et al. 2006) have been suggested that cybermetric measures could be useful for reflecting the contribution of technologically oriented institutions. High correlations between the size of the websites and the number of the files have been founded by them. Unfortunately, correlations between traffic of the websites and size of the websites and number of the files have not been considered by them.

Performance of the websites for having more visitors and keeping loyalty of them depends on how to provide their content. Recently, a lot of research is being carried out the field of analyzing and understanding user behavior on a Web site. Collecting number of the visitors and visit duration have been investigated for developing performance indicators by (Ortega & Aguillo, 2010). New indicator which can be used by research institutes for improved search collaboration has been introduced by (Lavhengwa, Lavhengwa & Jacobus, 2014), Web trust and web loyalty have been investigated by (Jamaludin, 2013) and (Weth, Christian & Hauswirth, 2014) are investigated browsing behavior of online users. Also, choosing the right tool for analyzing attraction of the visitors with a website have been explained by (Booth & Jansen , 2009) and a model for helping site's designer to understand behavior of site visitors that will be used to decide the human information behavior has been introduced by (Mishara, 2014).

Although, the relation between traffic of the websites with higher rated scholars production and more web content with average online impact has been shown by (Thelwall, Harries, 2004) but, they have not found evidence that higher rated scholars produce higher impact web content. Content-base studies of Google Scholar have been shown a close corresponding between the materials indexed by Google Scholar with those included in the Compendex database (Meier & Conkling, 2008), but Aguillo (2012) has suggested that the use of Google Scholar for evaluation purposes should be done with great care.

**Research Questions**

In this study, we address three questions to compare traffic of the academic websites and content of the websites.

-Is there a correlation between number of the papers, size of the websites and number of the files with traffic of the websites?

-Which one is the most important for attracting more visitors: papers, files or webpages?

-What are the rankings of countries based on correlation between academic contents and traffic of the websites?

**Methodology**

Traffic of the websites, size of the websites, number of the papers and number of the files are the four parameters which is used for finding visitors' interest in this study. For quantitative purposes, data of 21,485 universities and research centers have been extracted at first weak of the January, 2014.

For collecting data from the Alexa, we have used canonical address of Alexa (http://alexa.com) and the following syntax: *site:domain*. Google has been used for extracting size of the websites and number of the files. Syntax of the extracting size of the site in the Google is *site:domain*. Google provides the number of all file types (pdf, doc, ppt, docx, pptx, ps and xls), respectively. For collecting number of the each type we should define type of the files. For example for extracting the number of the pdf files in each site we use this syntax: *site:domain filetype:pdf*. Summation of the all file formats has been introduced as "Rich Files" by (Aguillo, et al, 2006). In this study, canonical address of Google Scholar (http://scholar.google.com) and the following syntax: *site:domain* have been used for extracting number of papers and citations.

Alexa provides rankings of the websites. Although, Alexa provide global rank of the websites, ranks of the universities have been distributed in the range of 1 to 30,000,000. In this study rank of the universities which haven't been provided by the Alexa assigned to 30,000,000. After collecting all data from Alexa, rankings of academic websites in each country reproduce. Ranking could be better answer for describing relative impact of an institutional web site (Aguillo, et al, 2006).

Google and Google Scholar provide number of papers, number of the files and size of the websites. Google and Google Scholar don't provide rankings of these parameters. For calculating correlation between Alexa's rankings and Google and Google Scholar data we have been produced rankings of these parameters. All rankings have been produced for each country, respectively.

At the end of this study we calculate correlation between Alexa, Google and Google Scholar for all universities. We investigate the visitor's interest in the world and compare with countries. We regenerate a list of all 21,485 universities and their ranks in the world.

In this study, largest set of the academic websites (universities and research institutes) in the World Wide Web has been analyzed. Our database contains 21,485 academic domains which have been covered by 197 countries and 1 international category (5 universities).

## Results

Table (1) shows the result of top 15 universities in the United State of America and their Alexa's global rank. Rankings of websites in United State of America which is called "Country-Alexa Rank" have been regenerated based on rankings of Alexa in the left column.

Rank of the Google Scholar has been generated based on their number of the papers. Table (2) shows the top 15 universities with highest number of the papers in the United State of America and their Country-Scholar Rank. These rankings have been generated for calculating correlation with Country-Alexa Rank.

Table 1

*Generating rankings of the websites in the United State of America based on their global rank in the Alexa*

| University Name | Domain | Alexa global rank | Country-Alexa Rank |
|---|---|---|---|
| Harvard-MIT Division of Health Sciences and Tecnology | hst.mit.edu | 1166 | 1 |
| Massachusetts Institute of Technology | mit.edu | 1174 | 2 |
| Stanford University | stanford.edu | 1400 | 3 |
| Harvard University | harvard.edu | 1553 | 4 |

| University Name | Domain | Alexa global rank | Country-Alexa Rank |
|---|---|---|---|
| University of California Berkeley | berkeley.edu | 2158 | 5 |
| Pennsylvania State University | psu.edu | 2385 | 6 |
| Columbia University New York | columbia.edu | 2679 | 7 |
| Cornell University | cornell.edu | 2890 | 8 |
| Weill Medical College Cornell University | med.cornell.edu | 2933 | 9 |
| University of Texas Austin | utexas.edu | 3188 | 10 |
| University of Michigan | umich.edu | 3307 | 11 |
| New York University | nyu.edu | 3330 | 12 |
| University of Michigan Dearborn | umd.umich.edu | 3359 | 13 |
| University of Wisconsin Madison | wisc.edu | 3876 | 14 |
| University of Pennsylvania | upenn.edu | 3945 | 15 |

Two additional rankings which have been generated in this study are rankings of websites based on the number of rich files and rankings of websites based on the size of the websites. Table (3) shows the result of top 50 universities and their ranks in the United State of America which has been sorted based on traffic of the websites. Rankings of the size, rankings of the rich files and rankings of the papers have been concluded in the right columns in table (3).

Table 2

*Generating rankings of the websites in the United State of America based on their papers number in the Google Scholar*

| University Name | Domain | Number of Papers | Country-Scholar Rank |
|---|---|---|---|
| Harvard University | harvard.edu | 2100000 | 1 |
| Pennsylvania State University | psu.edu | 371000 | 2 |
| Johns Hopkins University | jhu.edu | 256000 | 3 |
| Massachusetts Institute of Technology | mit.edu | 78500 | 4 |
| University of Minnesota | umn.edu | 52500 | 5 |
| Georgia Institute of Technology | gatech.edu | 46200 | 6 |
| Stanford University | stanford.edu | 45500 | 7 |
| University of Nebraska Lincoln | unl.edu | 42900 | 8 |
| Purdue University | purdue.edu | 38900 | 9 |
| University of Pittsburgh | pitt.edu | 38100 | 10 |
| University of Michigan | umich.edu | 38000 | 11 |
| Oregon State University | oregonstate.edu | 36400 | 12 |
| Cornell University | cornell.edu | 36000 | 13 |
| University of Texas Austin | utexas.edu | 33600 | 14 |
| Ohio State University | osu.edu | 33200 | 15 |

Table (3) has 7 columns. Data of column (3) are global ranks which have been extracted based on domain addresses in column (2). Ranks of columns 4, 5, 6 and 7 have been generated based on the real data. In each country, best rank is one and worth rank is equal to the number of domains in the country. For the best evaluation, all data have been extracted at January, 2014.

For measuring correlation between traffic of each site and three main parts of the site (Size, Rich Files and Papers), it's enough to calculate correlation between data of column (4) and columns 5, 6 and 7, respectively. Country-Alexa Rank, Country-Scholar Rank, Country-Size Rank and Country-Rich-Files Rank are parameters which scholars can use for calculating correlation between traffic of the websites and main parts of the websites.

Comparing correlation data in one country could help web designers to make better websites for increasing visitor's loyalty. If traffic-scholar correlation (correlation between column (4) and column (5)) in one country be greater than traffic-size correlation (correlation between column (4) and column (6)), websites could provide more papers for attracting more visitors. On the other hand, if the traffic-size correlation in one country be greater than traffic-scholar correlation, then its better for websites to provide more webpages and popular content for attracting more visitors.

At this stage the correlations between countries can be computed. For example, if the traffic-size correlation in first country is greater than traffic-size correlation in second country, performance of the academic websites in first country for obtaining more visitors is greater than performance of second country. If the traffic-scholar correlation in one country is greater than same correlation in second country, the impact of papers for attracting visitors in the first country is greater than performance of the second country.

Table 3

*Generating rankings of the top 50 websites in the United State of America and sorting on the Country-Alexa rank based on their global rank in the Alexa and data of websites in the Google and Google Scholar.*

| (1) University name | (2) Domain | (3) Alexa Global Rank | (4) Country-Alexa Rank | (5) Country-Scholar Rank | (6) Country-Size Rank | (7) Country-Rich-Files Rank |
|---|---|---|---|---|---|---|
| Harvard-MIT Division of Health Sciences and Technology | hst.mit.edu | 1166 | 1 | 1548 | 2630 | 2693 |
| Massachusetts Institute of Technology | mit.edu | 1174 | 2 | 4 | 24 | 5 |
| Stanford University | stanford.edu | 1400 | 3 | 7 | 36 | 10 |
| Harvard University | harvard.edu | 1553 | 4 | 1 | 3 | 13 |
| University of California Berkeley | berkeley.edu | 2158 | 5 | 20 | 70 | 14 |
| Pennsylvania State University | psu.edu | 2385 | 6 | 2 | 22 | 2 |
| Columbia University New York | columbia.edu | 2679 | 7 | 19 | 106 | 32 |
| Cornell University | cornell.edu | 2890 | 8 | 13 | 51 | 12 |
| Weill Medical College Cornell University | med.cornell.edu | 2933 | 9 | 695 | 863 | 1474 |
| University of Texas Austin | utexas.edu | 3188 | 10 | 14 | 104 | 19 |
| University of Michigan | umich.edu | 3307 | 11 | 11 | 63 | 15 |
| New York University | nyu.edu | 3330 | 12 | 49 | 9 | 50 |
| University of Michigan Dearborn | umd.umich.edu | 3359 | 13 | 513 | 562 | 556 |

| (1) University name | (2) Domain | (3) Alexa Global Rank | (4) Country-Alexa Rank | (5) Country-Scholar Rank | (6) Country-Size Rank | (7) Country-Rich-Files Rank |
|---|---|---|---|---|---|---|
| University of Wisconsin Madison | wisc.edu | 3876 | 14 | 25 | 66 | 9 |
| University of Pennsylvania | upenn.edu | 3945 | 15 | 18 | 92 | 48 |
| University of Minnesota | umn.edu | 3982 | 16 | 5 | 86 | 11 |
| University of Washington | washington.edu | 4004 | 17 | 21 | 81 | 4 |
| University of Minnesota Duluth | d.umn.edu | 4022 | 18 | 294 | 249 | 177 |
| University of Minnesota Morris | morris.umn.edu | 4042 | 19 | 529 | 773 | 628 |
| University of Minnesota Crookston | crk.umn.edu | 4051 | 20 | 2178 | 1293 | 770 |
| University of Minnesota, Rochester | r.umn.edu | 4067 | 21 | 2401 | 2509 | 2312 |
| University of Illinois Urbana Champaign | uiuc.edu | 4073 | 22 | 62 | 47 | 87 |
| University of California Los Angeles UCLA | ucla.edu | 4103 | 23 | 48 | 7 | 29 |
| Purdue University | purdue.edu | 4108 | 24 | 9 | 28 | 24 |
| Princeton University | princeton.edu | 4109 | 25 | 63 | 5 | 52 |
| University of Illinois | illinois.edu | 4142 | 26 | 22 | 84 | 20 |
| CUNY Medgar Evers College | mec.cuny.edu | 4600 | 27 | 2523 | 2430 | 1741 |
| CUNY John Jay College of Criminal Justice | jjay.cuny.edu | 4603 | 28 | 696 | 662 | 635 |
| City University of New York | cuny.edu | 4607 | 29 | 122 | 1 | 46 |
| CUNY New York City College of Technology | citytech.cuny.edu | 4610 | 30 | 581 | 461 | 527 |
| CUNY York College | york.cuny.edu | 4610 | 31 | 2817 | 695 | 2021 |
| CUNY Brooklyn College | brooklyn.cuny.edu | 4613 | 32 | 449 | 417 | 369 |
| CUNY Queens College | qc.cuny.edu | 4613 | 33 | 444 | 311 | 402 |
| City College of New York CUNY | ccny.cuny.edu | 4624 | 34 | 328 | 468 | 329 |
| CUNY Hunter College | hunter.cuny.edu | 4624 | 35 | 426 | 330 | 338 |
| CUNY College of Staten Island | csi.cuny.edu | 4629 | 36 | 672 | 974 | 753 |
| CUNY Baruch College | baruch.cuny.edu | 4629 | 37 | 459 | 260 | 433 |
| Yale University | yale.edu | 4679 | 38 | 47 | 118 | 53 |
| Carnegie Mellon University | cmu.edu | 4836 | 39 | 24 | 120 | 16 |
| University of Florida | ufl.edu | 5074 | 40 | 32 | 116 | 8 |
| University of Southern California | usc.edu | 5835 | 41 | 64 | 26 | 37 |
| Harvard University Harvard Business School | hbs.edu | 6132 | 42 | 268 | 285 | 573 |
| Ohio State University | osu.edu | 6220 | 43 | 15 | 16 | 33 |
| University of California Davis | ucdavis.edu | 6409 | 44 | 70 | 35 | 40 |
| University of California San Diego | ucsd.edu | 6736 | 45 | 67 | 6 | 21 |
| Rutgers University | rutgers.edu | 6752 | 46 | 39 | 2 | 22 |
| University of Phoenix | phoenix.edu | 6783 | 47 | 1489 | 901 | 1779 |
| Michigan State University | msu.edu | 6819 | 48 | 43 | 90 | 6 |
| Rutgers University Camden | camden.rutgers.edu | 6954 | 49 | 759 | 864 | 907 |
| University of Maryland | umd.edu | 7287 | 50 | 26 | 112 | 26 |

All countries which have at least one academic website are included in this study. Table (4) shows the top 40 countries with highest number of the active universities. Correlations between traffic of the websites in each country and major parts of the websites have been calculated.

World's correlation or visitor's interest in the world is acceptable by measuring correlation between traffic and main parts of the websites for all universities in one category. For calculating correlation between traffic of the websites and their content in the world, we have repeated the method of generating rank for all the 21,485 websites in one category. In this case, we can compare the visitors' interest in the each country with the visitors' interest in the world. First row in table (4) belongs to world's correlations between traffic of the 21,485 websites and their three main parts of their contents.

Table (4) shows that the Alexa-Size correlation is similar to the Alexa-Rich-Files correlation. Similarity between the Alexa-Size and the Alexa-Rich-Files correlation is due to strategy of the Google for creating report of the websites. In the Google, size of the websites is equal to summation of number of the pages and number of the rich files, therefore the number of the rich files has been included in the size of the websites. Correlations are calculated based on Spearman method.

As mentioned earlier, rich files is the summation of all file types (pdf, doc, docx, ppt, pptx, xls and ps). Most files are in the pdf, doc and ppt format and the number of pptx, xls and ps files which have been extracted are not considerable. We added three columns for investigating correlation between traffic of the websites with these file types.

Similarity between Alexa-Rich-Files and Alexa-PDF correlations in table (4) shows that most of the files which have been created in the websites are in pdf format and visitors are interested in using pdf file formats. Using of pdf file format is more than using of other types of the files format in most of countries.

Data of the world in table (4) shows that correlation between traffic of the websites and the size of the websites and rich files are equal to 0.71. Although, results show that for obtaining more visitors in the world, creating of rich files and webpages have same effect. With combining results of correlation between worlds and each country we lead to make more webpages for obtaining more local visitors and creating more pdf files for obtaining global visitors around the world.

In the world and all of the top countries, Spearman correlation between traffic and size of the websites is greater than the correlation between traffic and the number of the papers. Due to this result, for attracting more visitors, the effect of creating more papers is less than effect of creating more webpages.

Table 4

*Shows the correlation between the traffic of the websites and the Size, Rich file and papers in the top 40 countries with highest number of the active universities.*

|  | Country Name | Number of Universities | Alexa-Size Correlation | Alexa-Rich-Files Correlation | Alexa-Scholar Correlation | Alexa-PDF Correlation | Alexa-DOC Correlation | Alexa-PPT Correlation |
|---|---|---|---|---|---|---|---|---|
|  | World | 21,485 | 0.71 | 0.71 | 0.61 | 0.72 | 0.62 | 0.61 |
| 1 | United States of America | 3344 | 0.75 | 0.70 | 0.59 | 0.70 | 0.66 | 0.63 |
| 2 | Brazil | 1834 | 0.70 | 0.65 | 0.58 | 0.66 | 0.57 | 0.53 |
| 3 | India | 1743 | 0.77 | 0.75 | 0.50 | 0.75 | 0.54 | 0.51 |

| | Country Name | Number of Universities | Alexa-Size Correlation | Alexa-Rich-Files Correlation | Alexa-Scholar Correlation | Alexa-PDF Correlation | Alexa-DOC Correlation | Alexa-PPT Correlation |
|---|---|---|---|---|---|---|---|---|
| 4 | China | 1252 | 0.80 | 0.79 | 0.68 | 0.78 | 0.69 | 0.64 |
| 5 | Russian Federation | 1088 | 0.71 | 0.62 | 0.52 | 0.63 | 0.56 | 0.56 |
| 6 | Mexico | 962 | 0.61 | 0.52 | 0.46 | 0.51 | 0.49 | 0.45 |
| 7 | Japan | 861 | 0.71 | 0.66 | 0.55 | 0.65 | 0.62 | 0.57 |
| 8 | France | 635 | 0.64 | 0.63 | 0.55 | 0.63 | 0.59 | 0.55 |
| 9 | Iran (Islamic Republic of Iran) | 605 | 0.80 | 0.70 | 0.53 | 0.70 | 0.61 | 0.52 |
| 10 | Poland | 475 | 0.70 | 0.66 | 0.57 | 0.65 | 0.61 | 0.56 |
| 11 | Germany | 425 | 0.82 | 0.81 | 0.71 | 0.81 | 0.76 | 0.66 |
| 12 | Republic Of Korea | 419 | 0.59 | 0.59 | 0.51 | 0.59 | 0.58 | 0.56 |
| 13 | Indonesia | 373 | 0.85 | 0.81 | 0.70 | 0.80 | 0.75 | 0.68 |
| 14 | Pakistan | 344 | 0.78 | 0.74 | 0.61 | 0.72 | 0.64 | 0.58 |
| 15 | Ukraine | 336 | 0.67 | 0.63 | 0.55 | 0.62 | 0.55 | 0.56 |
| 16 | United Kingdom | 330 | 0.81 | 0.80 | 0.73 | 0.80 | 0.76 | 0.74 |
| 17 | Philippines | 307 | 0.49 | 0.49 | 0.48 | 0.50 | 0.46 | 0.50 |
| 18 | Colombia | 306 | 0.81 | 0.77 | 0.59 | 0.77 | 0.65 | 0.64 |
| 19 | Canada | 265 | 0.77 | 0.74 | 0.66 | 0.74 | 0.70 | 0.67 |
| 20 | Spain | 248 | 0.91 | 0.80 | 0.71 | 0.80 | 0.66 | 0.71 |
| 21 | Italy | 225 | 0.83 | 0.77 | 0.73 | 0.77 | 0.69 | 0.63 |
| 22 | Thailand | 183 | 0.83 | 0.84 | 0.67 | 0.85 | 0.79 | 0.75 |
| 23 | Turkey | 170 | 0.85 | 0.80 | 0.76 | 0.80 | 0.72 | 0.75 |
| 24 | Taiwan | 170 | 0.81 | 0.80 | 0.66 | 0.80 | 0.76 | 0.74 |
| 25 | Netherlands | 156 | 0.87 | 0.73 | 0.61 | 0.74 | 0.56 | 0.62 |
| 26 | Nigeria | 144 | 0.71 | 0.71 | 0.33 | 0.72 | 0.40 | 0.29 |
| 27 | Vietnam | 124 | 0.80 | 0.71 | 0.58 | 0.69 | 0.68 | 0.55 |
| 28 | Kazakstan | 120 | 0.66 | 0.60 | 0.39 | 0.55 | 0.56 | 0.50 |
| 29 | Portugal | 118 | 0.72 | 0.68 | 0.63 | 0.68 | 0.65 | 0.64 |
| 30 | Argentina | 117 | 0.75 | 0.67 | 0.63 | 0.67 | 0.60 | 0.59 |
| 31 | Switzerland | 113 | 0.78 | 0.73 | 0.58 | 0.74 | 0.63 | 0.58 |
| 32 | Romania | 111 | 0.85 | 0.84 | 0.68 | 0.85 | 0.73 | 0.62 |
| 33 | Bangladesh | 107 | 0.70 | 0.73 | 0.50 | 0.73 | 0.53 | 0.46 |
| 34 | Morocco | 105 | 0.56 | 0.62 | 0.26 | 0.60 | 0.51 | 0.40 |
| 35 | Australia | 104 | 0.74 | 0.67 | 0.63 | 0.67 | 0.64 | 0.66 |
| 36 | Belgium | 99 | 0.78 | 0.70 | 0.55 | 0.69 | 0.63 | 0.62 |
| 37 | Denmark | 98 | 0.76 | 0.70 | 0.48 | 0.70 | 0.65 | 0.58 |
| 38 | Peru | 92 | 0.75 | 0.60 | 0.59 | 0.62 | 0.50 | 0.40 |
| 39 | Chile | 85 | 0.76 | 0.78 | 0.54 | 0.78 | 0.65 | 0.62 |
| 40 | Czech Republic | 85 | 0.70 | 0.69 | 0.61 | 0.70 | 0.61 | 0.52 |

Table 5

*Sorting countries based on the best performance between traffic and size of the websites*

|  | Country Name | Number of Universities | Alexa-Size Correlation | Alexa-Rich-Files Correlation | Alexa-Scholar Correlation | Alexa-PDF Correlation | Alexa-DOC Correlation | Alexa-PPT Correlation |
|---|---|---|---|---|---|---|---|---|
| 1 | Spain | 248 | 0.91 | 0.80 | 0.71 | 0.80 | 0.66 | 0.71 |
| 2 | Netherlands | 156 | 0.87 | 0.73 | 0.61 | 0.74 | 0.56 | 0.62 |
| 3 | Turkey | 170 | 0.85 | 0.80 | 0.76 | 0.80 | 0.72 | 0.75 |
| 4 | Romania | 111 | 0.85 | 0.84 | 0.68 | 0.85 | 0.73 | 0.62 |
| 5 | Indonesia | 373 | 0.85 | 0.81 | 0.70 | 0.80 | 0.75 | 0.68 |
| 6 | Thailand | 183 | 0.83 | 0.84 | 0.67 | 0.85 | 0.79 | 0.75 |
| 7 | Italy | 225 | 0.83 | 0.77 | 0.73 | 0.77 | 0.69 | 0.63 |
| 8 | Germany | 425 | 0.82 | 0.81 | 0.71 | 0.81 | 0.76 | 0.66 |
| 9 | Colombia | 306 | 0.81 | 0.77 | 0.59 | 0.77 | 0.65 | 0.64 |
| 10 | Taiwan | 170 | 0.81 | 0.80 | 0.66 | 0.80 | 0.76 | 0.74 |
| 11 | United Kingdom | 330 | 0.81 | 0.80 | 0.73 | 0.80 | 0.76 | 0.74 |
| 12 | Iran (Islamic Republic of Iran) | 605 | 0.80 | 0.70 | 0.53 | 0.70 | 0.61 | 0.52 |
| 13 | Vietnam | 124 | 0.80 | 0.71 | 0.58 | 0.69 | 0.68 | 0.55 |
| 14 | China | 1252 | 0.80 | 0.79 | 0.68 | 0.78 | 0.69 | 0.64 |
| 15 | Belgium | 99 | 0.78 | 0.70 | 0.55 | 0.69 | 0.63 | 0.62 |
| 16 | Switzerland | 113 | 0.78 | 0.73 | 0.58 | 0.74 | 0.63 | 0.58 |
| 17 | Pakistan | 344 | 0.78 | 0.74 | 0.61 | 0.72 | 0.64 | 0.58 |
| 18 | India | 1743 | 0.77 | 0.75 | 0.50 | 0.75 | 0.54 | 0.51 |
| 19 | Canada | 265 | 0.77 | 0.74 | 0.66 | 0.74 | 0.70 | 0.67 |
| 20 | Denmark | 98 | 0.76 | 0.70 | 0.48 | 0.70 | 0.65 | 0.58 |
| 21 | Chile | 85 | 0.76 | 0.78 | 0.54 | 0.78 | 0.65 | 0.62 |
| 22 | Peru | 92 | 0.75 | 0.60 | 0.59 | 0.62 | 0.50 | 0.40 |
| 23 | United States of America | 3344 | 0.75 | 0.70 | 0.59 | 0.70 | 0.66 | 0.63 |
| 24 | Argentina | 117 | 0.75 | 0.67 | 0.63 | 0.67 | 0.60 | 0.59 |
| 25 | Australia | 104 | 0.74 | 0.67 | 0.63 | 0.67 | 0.64 | 0.66 |
| 26 | Portugal | 118 | 0.72 | 0.68 | 0.63 | 0.68 | 0.65 | 0.64 |
| 27 | Nigeria | 144 | 0.71 | 0.71 | 0.33 | 0.72 | 0.40 | 0.29 |
| 28 | Russian Federation | 1088 | 0.71 | 0.62 | 0.52 | 0.63 | 0.56 | 0.56 |
|  | World | 21,485 | 0.71 | 0.71 | 0.61 | 0.72 | 0.62 | 0.61 |
| 29 | Japan | 861 | 0.71 | 0.66 | 0.55 | 0.65 | 0.62 | 0.57 |
| 30 | Bangladesh | 107 | 0.70 | 0.73 | 0.50 | 0.73 | 0.53 | 0.46 |
| 31 | Brazil | 1834 | 0.70 | 0.65 | 0.58 | 0.66 | 0.57 | 0.53 |
| 32 | Czech Republic | 85 | 0.70 | 0.69 | 0.61 | 0.70 | 0.61 | 0.52 |
| 33 | Poland | 475 | 0.70 | 0.66 | 0.57 | 0.65 | 0.61 | 0.56 |
| 34 | Ukraine | 336 | 0.67 | 0.63 | 0.55 | 0.62 | 0.55 | 0.56 |
| 35 | Kazakstan | 120 | 0.66 | 0.60 | 0.39 | 0.55 | 0.56 | 0.50 |
| 36 | France | 635 | 0.64 | 0.63 | 0.55 | 0.63 | 0.59 | 0.55 |
| 37 | Mexico | 962 | 0.61 | 0.52 | 0.46 | 0.51 | 0.49 | 0.45 |
| 38 | Republic Of Korea | 419 | 0.59 | 0.59 | 0.51 | 0.59 | 0.58 | 0.56 |
| 39 | Morocco | 105 | 0.56 | 0.62 | 0.26 | 0.60 | 0.51 | 0.40 |
| 40 | Philippines | 307 | 0.49 | 0.49 | 0.48 | 0.50 | 0.46 | 0.50 |

Table (5) has been sorted based on the Alexa-Size correlation. Table (5) shows that maximum performance between the Alexa and size of the websites belongs to Spain and minimum performance belongs to Philippines.

Table (6) which have been sorted on Alexa-Scholar correlation, shows that the maximum performance between the Alexa and number of the papers papers belongs to Turkey and minimum performance belongs to Morocco.

Table 6
*Sorting countries based on the best performance between traffic of the websites and the number of the papers*

|  | Country Name | Number of Universities | Alexa-Size correlation | Alexa-Rich correlation | Alexa-Scholar correlation |
|---|---|---|---|---|---|
| 1 | Turkey | 170 | 0.85 | 0.80 | 0.76 |
| 2 | United Kingdom | 330 | 0.81 | 0.80 | 0.73 |
| 3 | Italy | 225 | 0.83 | 0.77 | 0.73 |
| 4 | Spain | 248 | 0.91 | 0.80 | 0.71 |
| 5 | Germany | 425 | 0.82 | 0.81 | 0.71 |
| 6 | Indonesia | 373 | 0.85 | 0.81 | 0.70 |
| 7 | Romania | 111 | 0.85 | 0.84 | 0.68 |
| 8 | China | 1252 | 0.80 | 0.79 | 0.68 |
| 9 | Thailand | 183 | 0.83 | 0.84 | 0.67 |
| 10 | Taiwan | 170 | 0.81 | 0.80 | 0.66 |
| 11 | Canada | 265 | 0.77 | 0.74 | 0.66 |
| 12 | Australia | 104 | 0.74 | 0.67 | 0.63 |
| 13 | Argentina | 117 | 0.75 | 0.67 | 0.63 |
| 14 | Portugal | 118 | 0.72 | 0.68 | 0.63 |
| 15 | Pakistan | 344 | 0.78 | 0.74 | 0.61 |
| 16 | Netherlands | 156 | 0.87 | 0.73 | 0.61 |
|  | World | 21,485 | 0.71 | 0.71 | 0.61 |
| 17 | Czech Republic | 85 | 0.70 | 0.69 | 0.61 |
| 18 | Colombia | 306 | 0.81 | 0.77 | 0.59 |
| 19 | Peru | 92 | 0.75 | 0.60 | 0.59 |
| 20 | United States of America | 3344 | 0.75 | 0.70 | 0.59 |
| 21 | Switzerland | 113 | 0.78 | 0.73 | 0.58 |
| 22 | Vietnam | 124 | 0.80 | 0.71 | 0.58 |
| 23 | Brazil | 1834 | 0.70 | 0.65 | 0.58 |
| 24 | Poland | 475 | 0.70 | 0.66 | 0.57 |
| 25 | France | 635 | 0.64 | 0.63 | 0.55 |
| 26 | Japan | 861 | 0.71 | 0.66 | 0.55 |
| 27 | Belgium | 99 | 0.78 | 0.70 | 0.55 |
| 28 | Ukraine | 336 | 0.67 | 0.63 | 0.55 |
| 29 | Chile | 85 | 0.76 | 0.78 | 0.54 |
| 30 | Iran (Islamic Republic of Iran) | 605 | 0.80 | 0.70 | 0.53 |

|    | Country Name | Number of Universities | Alexa-Size correlation | Alexa-Rich correlation | Alexa-Scholar correlation |
|----|---|---|---|---|---|
| 31 | Russian Federation | 1088 | 0.71 | 0.62 | 0.52 |
| 32 | Republic Of Korea | 419 | 0.59 | 0.59 | 0.51 |
| 33 | India | 1743 | 0.77 | 0.75 | 0.50 |
| 34 | Bangladesh | 107 | 0.70 | 0.73 | 0.50 |
| 35 | Philippines | 307 | 0.49 | 0.49 | 0.48 |
| 36 | Denmark | 98 | 0.76 | 0.70 | 0.48 |
| 37 | Mexico | 962 | 0.61 | 0.52 | 0.46 |
| 38 | Kazakhstan | 120 | 0.66 | 0.60 | 0.39 |
| 39 | Nigeria | 144 | 0.71 | 0.71 | 0.33 |
| 40 | Morocco | 105 | 0.56 | 0.62 | 0.26 |

Result show that although, there is a significant correlation between traffic of the websites and number of the papers, for obtaining more traffic, performance of the papers are less than performance of the files and pages.

**Discussion**

The study found that we can compare performance of the academic websites by computing the correlation between traffic and content of the websites. This finding is in agreement with previous study (Vaughan & Yang, 2013) which showed that academic quality performance has relation with traffic of the websites. This achievement is in the same trend of findings from earlier studies (Aguillo, et al, 2006; Thelwall & Sud, 2011).

Characteristics of the Alexa are our main limitations. Alexa provides only the last three months traffic for each site. Data of Alexa is gathered from computers which Alexa's toolbar is installed on them. Our extracted data shows that differences between traffic ranks of the main domains to their subdomains are very low. Alexa assigns traffic ranks of the main domains to their subdomains with a little difference.

The sample time span is another limitation. Most of the websites' visitors are students of universities. Different vacation and registration time in all countries lead to different traffics at the extracting time.

As mentioned earlier, we can introduce best countries and rankings of countries based on the performance of the each structure (Size, Papers or Rich Files) for attracting more visitors. Although, we can introduce rankings of countries based on the performance of the each part of the academic websites, but still, we cannot introduce best countries in the performance of academic websites. We have also investigated two complex formula based on performance of the each part.

Table (7) shows that all data in the column (Alexa-Size)-(Alexa-Scholar) are greater than zero which means that in all of the 40 countries, correlation between traffic of the websites and size of the websites is greater than the correlation between traffic of the websites and the number of papers. Therefore, in the entire world, site's designer should create more webpages than papers for increasing traffic of the websites. The column (Alexa-Rich)-(Alexa-Scholar) shows that interest of visitors for using rich files are more than using papers in the world.

For finding the reason of these results, number of the papers in the all universities has been investigated in this study. Our database which all of its data have been extracted in the first week of the January, 2014, shows that probability of accepting files and pages in the Google is more than probability of accepting papers in the Google Scholar. The number of papers in 7,281 universities of all universities in the world is more than zero and 14,204 academic websites didn't have any paper in the Google Scholar. Google Scholar has rigid standards for accepting papers. Although, some repositories like Dbspace and Eprints have been introduced by Google Scholar for helping universities to register their papers, but, the number of papers in 14,204 academic websites in the Google Scholar was equal to zero. On the other hand, Google has simple rules for accepting files and pages. Our database shows that only 764 websites had zero size and the 2535 websites had no rich files.

For summarizing the results of the three correlations to one indicator, we have suggested the multiplication of these three correlations. If we calculate the multiplication of all correlations in each country, we have a new indicator which indicates power of each country for attracting more visitors depend on the performance of each part. This method can introduce one indicator for summarizing all results of traffic correlations.

Introducing a new indicator based on the multiplication of correlations instead of summation of them, help scholars for investigating equality of these parameters. In this case, if two countries have a same summation of performance, then country which has more multiplication is more reliable.

Table (7) shows the result of multiplication of three correlations in each country. All countries are sorted based on (Alexa-Size)*(Alexa-Rich)*(Alexa-Scholar).

There is no correlation between the number of the universities in the countries and performance of the websites. We have used Spearman method for calculating the correlation between the number of the universities in each country and the Alexa-Size correlation. Calculated correlation was negative (-0.08) and a bit less than zero.

Table 7

*Sorting countries based on the multiplication of all correlations*

| | Country Name | Number of Universities | AlexaSize | AlexaRich | AlexaScholar | (Alexa-Size)*(Alexa-Rich)*(Alexa-Scholar) | (Alexa-Size)-(Alexa-Rich) | (Alexa-Size)-(Alexa-Scholar) | (Alexa-Rich)-(Alexa-Scholar) |
|---|---|---|---|---|---|---|---|---|---|
| 1 | Turkey | 170 | 0.85 | 0.80 | 0.76 | 0.52 | 0.05 | 0.09 | 0.04 |
| 2 | Spain | 248 | 0.91 | 0.80 | 0.71 | 0.52 | 0.11 | 0.19 | 0.09 |
| 3 | Romania | 111 | 0.85 | 0.84 | 0.68 | 0.49 | 0.01 | 0.17 | 0.16 |
| 4 | Indonesia | 373 | 0.85 | 0.81 | 0.70 | 0.48 | 0.04 | 0.14 | 0.10 |
| 5 | United Kingdom | 330 | 0.81 | 0.80 | 0.73 | 0.47 | 0.00 | 0.08 | 0.07 |
| 6 | Thailand | 183 | 0.83 | 0.84 | 0.67 | 0.47 | -0.01 | 0.16 | 0.17 |
| 7 | Germany | 425 | 0.82 | 0.81 | 0.71 | 0.47 | 0.02 | 0.12 | 0.10 |
| 8 | Italy | 225 | 0.83 | 0.77 | 0.73 | 0.46 | 0.06 | 0.10 | 0.04 |
| 9 | Taiwan | 170 | 0.81 | 0.80 | 0.66 | 0.43 | 0.01 | 0.15 | 0.13 |

| | Country Name | Number of Universities | AlexaSize | AlexaRich | AlexaScholar | (Alexa-Size)*(Alexa-Rich)*(Alexa-Scholar) | (Alexa-Size)-(Alexa-Rich) | (Alexa-Size)-(Alexa-Scholar) | (Alexa-Rich)-(Alexa-Scholar) |
|---|---|---|---|---|---|---|---|---|---|
| 10 | China | 1252 | 0.80 | 0.79 | 0.68 | 0.43 | 0.01 | 0.12 | 0.11 |
| 11 | Netherlands | 156 | 0.87 | 0.73 | 0.61 | 0.39 | 0.14 | 0.26 | 0.12 |
| 12 | Canada | 265 | 0.77 | 0.74 | 0.66 | 0.38 | 0.03 | 0.11 | 0.08 |
| 13 | Colombia | 306 | 0.81 | 0.77 | 0.59 | 0.37 | 0.05 | 0.22 | 0.17 |
| 14 | Pakistan | 344 | 0.78 | 0.74 | 0.61 | 0.35 | 0.03 | 0.16 | 0.13 |
| 15 | Switzerland | 113 | 0.78 | 0.73 | 0.58 | 0.33 | 0.05 | 0.20 | 0.15 |
| 16 | Vietnam | 124 | 0.80 | 0.71 | 0.58 | 0.33 | 0.09 | 0.22 | 0.13 |
| 17 | Argentina | 117 | 0.75 | 0.67 | 0.63 | 0.32 | 0.07 | 0.12 | 0.04 |
| 18 | Chile | 85 | 0.76 | 0.78 | 0.54 | 0.32 | -0.02 | 0.22 | 0.24 |
| 19 | United States of America | 3344 | 0.75 | 0.70 | 0.59 | 0.31 | 0.05 | 0.16 | 0.11 |
| 20 | Australia | 104 | 0.74 | 0.67 | 0.63 | 0.31 | 0.07 | 0.10 | 0.03 |
| 21 | Portugal | 118 | 0.72 | 0.68 | 0.63 | 0.31 | 0.04 | 0.09 | 0.05 |
| | World | 21,485 | 0.71 | 0.71 | 0.61 | 0.31 | -0.01 | 0.10 | 0.11 |
| 22 | Belgium | 99 | 0.78 | 0.70 | 0.55 | 0.30 | 0.09 | 0.24 | 0.15 |
| 23 | Iran (Islamic Republic of Iran) | 605 | 0.80 | 0.70 | 0.53 | 0.29 | 0.10 | 0.27 | 0.18 |
| 24 | India | 1743 | 0.77 | 0.75 | 0.50 | 0.29 | 0.02 | 0.27 | 0.25 |
| 25 | Czech Republic | 85 | 0.70 | 0.69 | 0.61 | 0.29 | 0.01 | 0.09 | 0.09 |
| 26 | Peru | 92 | 0.75 | 0.60 | 0.59 | 0.27 | 0.15 | 0.16 | 0.01 |
| 27 | Poland | 475 | 0.70 | 0.66 | 0.57 | 0.26 | 0.04 | 0.13 | 0.09 |
| 28 | Brazil | 1834 | 0.70 | 0.65 | 0.58 | 0.26 | 0.05 | 0.12 | 0.08 |
| 29 | Bangladesh | 107 | 0.70 | 0.73 | 0.50 | 0.26 | -0.03 | 0.20 | 0.23 |
| 30 | Japan | 861 | 0.71 | 0.66 | 0.55 | 0.26 | 0.05 | 0.16 | 0.11 |
| 31 | Denmark | 98 | 0.76 | 0.70 | 0.48 | 0.25 | 0.06 | 0.28 | 0.23 |
| 32 | Ukraine | 336 | 0.67 | 0.63 | 0.55 | 0.23 | 0.04 | 0.13 | 0.08 |
| 33 | Russian Federation | 1088 | 0.71 | 0.62 | 0.52 | 0.23 | 0.09 | 0.19 | 0.10 |
| 34 | France | 635 | 0.64 | 0.63 | 0.55 | 0.22 | 0.01 | 0.09 | 0.08 |
| 35 | Republic Of Korea | 419 | 0.59 | 0.59 | 0.51 | 0.18 | -0.01 | 0.08 | 0.09 |
| 36 | Nigeria | 144 | 0.71 | 0.71 | 0.33 | 0.16 | -0.01 | 0.38 | 0.39 |
| 37 | Kazakstan | 120 | 0.66 | 0.60 | 0.39 | 0.15 | 0.06 | 0.27 | 0.21 |
| 38 | Mexico | 962 | 0.61 | 0.52 | 0.46 | 0.15 | 0.09 | 0.15 | 0.05 |
| 39 | Philippines | 307 | 0.49 | 0.49 | 0.48 | 0.12 | 0.00 | 0.01 | 0.01 |
| 40 | Morocco | 105 | 0.56 | 0.62 | 0.26 | 0.09 | -0.07 | 0.29 | 0.36 |

In response to the first research question, there were significant positive correlation between traffic of the academic websites and quantity of the Size, Rich Files and the Papers.

In answer to the second research question, there were a high correlation between the traffic of the websites and the size of the websites. Correlation between traffic of the websites and Rich Files is close to the correlation of the traffic of the websites and size of the websites.

Data shows that performance of the papers is lower than performance of the webpages and the files. Finally, rankings of countries based on performance of the countries for attracting more visitors are introduced in Table (5) and Table (6).

## Conclusion

We concluded that, there is positive correlation between traffic of the academic websites and size of the websites, number of the papers and number of the Rich Files (pdf, doc and ppt file formats). Performance of the academic websites for providing data and attracting more visitors in the web pages and Rich Files are greater than performance of the papers in the websites. Visitor's interest for using of the pdf files format is more than using doc and ppt file formats.

Google and Google Scholar are reliable for extracting data of the domains and subdomains. Although, the Alexa is a good source for extracting traffic of the main domains, we don't recommend for using the Alexa for subdomains.

It is recommended to investigate relation between country's academic traffic and format of the webpages (html, php and asp) and speed of the internet in the countries.